**Type-II magnetic-domain contrast caused by internal-magnetisation–induced electron-scattering anisotropy**

Tomohito Tanaka[1]*, Kazuto Kawakami[2], Hisashi Mogi[1], Satoshi Arai[2]

corresponding author: tanaka.m9p.tomohito@jp.nipponsteel.com

[1] Nippon Steel Corporation, 20-1 Shintomi, Futtsu City, Chiba Prefecture, JAPAN

[2] Nippon Steel Technology, 20-1 Shintomi, Futtsu City, Chiba Prefecture, JAPAN

**Abstract**

Previous studies have shown that type-II magnetic-domain contrasts are caused by differences in the backscattering yields of magnetic domains of opposite magnetisation. Imaging the magnetic domains when the magnetisation vectors in the opposite-magnetisation domains are perpendicular to the tilt axis of the specimen has been considered impossible, because of the lack of change in the backscattering yields between the domains. However, we found that it is possible to observe the type-II magnetic-domain contrast even in cases where the backscattering coefficients of the domains are the same. The contrast arises from the electron-scattering anisotropy created because of the deflection of the electron trajectories by the Lorentz force. We verify this by distinguishing all four possible in-plane magnetisation vectors on a Fe–Si (100) surface, using an electron backscatter diffraction detector as an array of electron detectors. The change in contrast between the magnetic domains, with respect to the location of a virtual electron detector, can provide information on the directions of the magnetisation vectors. A method to suppress the topographic contrast superimposed on the magnetic-domain contrast is also demonstrated.

## Introduction

Observation of magnetic domains helps us develop advanced magnetic materials. Among the characterisation techniques used for visualising magnetic domains, scanning electron microscopy (SEM) offers high spatial resolution [1]. Additional information such as elemental and crystallographic orientation distributions can also be acquired through SEM by using an energy-dispersive X-ray spectrometer (EDX) and an electron backscatter diffraction (EBSD) detector [1, 2].

Out of the three types of magnetic-domain contrasts that can appear in SEM images [1, 3-17], the type-II magnetic-domain contrast is suitable for visualising magnetic domains of ferromagnetic materials with cubic anisotropy, such as Fe–Si steel [1, 4, 8-15]. In general, a type-II magnetic-domain contrast is obtained by capturing a backscattered (or forward-scattered) electron image or by measuring the specimen current from a highly inclined specimen subjected to focused electron-beam irradiation. In the case of Fe–Si steel, the inclination angle should be between 50 and 60° to the horizontal, to maximise contrast [9]. Some researchers set an angle of 70° to satisfy the typical EBSD analysis configuration and use forward-scattered electron detectors to analyse the magnetic domains [12-15].

It is generally understood that the type-II contrast is generated simply because of a "number effect" [1, 8, 9, 18, 19]. The backscattering yields from domains of opposite magnetisations normally differ owing to the action of the Lorentz force. For example, the maximum possible Lorentz force acting on electrons as they enter a specimen prior to scattering is directed towards the surface in the upper domain and away from the surface in the lower domain, in **Fig. 1**(a). This results in a slightly higher backscattering yield from the upper-domain surface. Thus, the upper domain is brighter than the lower domain in the back (or forward-) scattered electron image.

It is also known that the type-II magnetic-domain contrast disappears when the magnetisation vectors in domains of opposite magnetisation are perpendicular to the tilt axis of the specimen (**Fig. 1**(b)) [1, 8, 9]. In this case, the direction of the Lorentz force acting on the electrons, as they enter the specimen prior to scattering, is parallel to the tilt axis, and no change occurs in the backscattering yields between the domains. It is necessary to rotate the specimen about the surface normal in order to obtain the magnetic-domain contrast.

In this study, we seek a way to obtain the type-II magnetic-domain contrast even in the case of **Fig. 1**(b), where the number of electrons emitted from the surface is (nearly) identical for domains of opposite magnetisation. The idea is very simple: if there exists a change in the anisotropy of the electron scattering caused by the Lorentz force, it should be possible to obtain the magnetic-domain contrast by utilising a position-sensitive electron detector. For instance, the magnetic domain with the magnetisation vector pointing upward (left magnetic domain in **Fig. 1**(b)) may be brighter in the scattered electron image, when the electron detector is divided into several portions and only the left portion of the detector is used. This is because the Lorentz force acting on the electrons prior to scattering points to the left in this magnetic domain; therefore, the electrons might tend to be scattered leftwards.

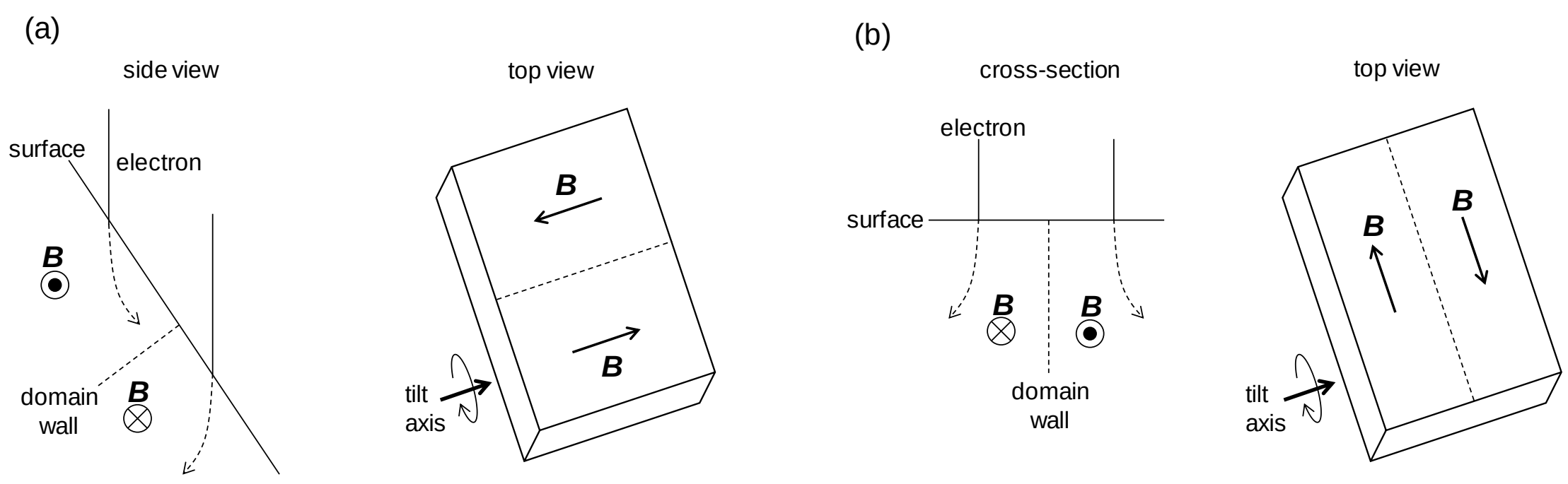

**Figure 1**. Schematic diagrams showing the change in electron trajectories (dashed arrows) due to the action of the Lorentz force when magnetisation vectors of opposite magnetisation are (a) parallel or (b) perpendicular to the tilt axis of the specimen. Solid arrows show the magnetisation vectors. Dashed lines depict the location of the magnetic-domain wall.

To verify this idea, we use an EBSD detector as an array of electron detectors [12, 20-22] and analyse the magnetic domains of a cube-textured Fe–Si (100)[001] surface. The distinction of all four possible in-plane magnetisation vectors ($B[100]$, $B[\bar{1}00]$, $B[010]$, $B[0\bar{1}0]$, where $B$ is the magnetic flux density) is attempted through the change in the magnetic-domain contrast with respect to the location of the electron detector.

A Monte Carlo electron-trajectory simulation incorporating magnetisation in an Fe specimen is also performed in this study. Previously, Monte Carlo simulations performed to explain the type-II magnetic-domain contrast [8, 18, 19] focused on alterations in the backscattering coefficients produced by magnetic domains with different magnetisation vectors. Our study considers the changes in the electron intensity distribution on an electron detector, as well.

Because the type-II magnetic domain contrast is known to be very small (in the order of a few tenths of one percent, under typical analysis conditions using a commercial SEM) [1, 4], topographic contrast often hinders the observation of magnetic-domain contrast. We also explore a route to suppress this topographic contrast superimposed on the magnetic-domain contrast, in this work.

## Methodology

### Materials

We prepared a cube-textured Fe-3 wt.% Si steel sheet by a cross-rolling method and secondary recrystallization [23, 24]. The rolling surface was nearly parallel to the (100) plane and the surface normal direction was nearly parallel to [001]. The specimen was cut to a size of 10 × 10 × 0.4 $mm^3$, and the surface was mechanically polished to a mirror finish. Electropolishing was then performed to remove the damaged layer near the surface, caused by the mechanical polishing.

### EBSD measurements and construction of electron images from EBSD patterns

The specimen was inclined at 70° from the horizontal, inside the SEM chamber. The acceleration voltage of the SEM (JEOL JSM-7001F) was set to 25 kV with a beam current of 14 nA. The working distance (WD) was set to 17 mm. All EBSD patterns were saved as 12-bit-depth tiff images. No background correction was performed. Each EBSD pattern was recorded with a resolution of 119 × 119 pixels, using a circular phosphor screen (EDAX DigiView camera). Although background correction was not applied, the patterns were indexable, as several Kikuchi bands were clearly visible. The EBSD crystal orientation map was obtained by scanning an electron beam over the surface. A step size of 1 μm, over a square grid, was used. Before the crystal-orientation analysis, the EBSD system was calibrated by matching the experimental EBSD pattern obtained from the centre of the EBSD map with the dynamically simulated EBSD pattern [25-27]. The experimental EBSD pattern was background-corrected only for calibration purposes. The crystal orientation was determined using the TSL OIM Data Collection software (EDAX).

We followed the method in [22] to create electron images from the stored EBSD patterns. The EBSD patterns were binned down to 5 × 5 superpixels (bins) using in-house MATLAB code on a computer, as schematically shown in **Fig. 2**. The intensities within the bins at each analysis point in the scan area were used to create a set of individual scattered electron images. All the electron images from $I_1$ to $I_{25}$ are presented in the Appendix.

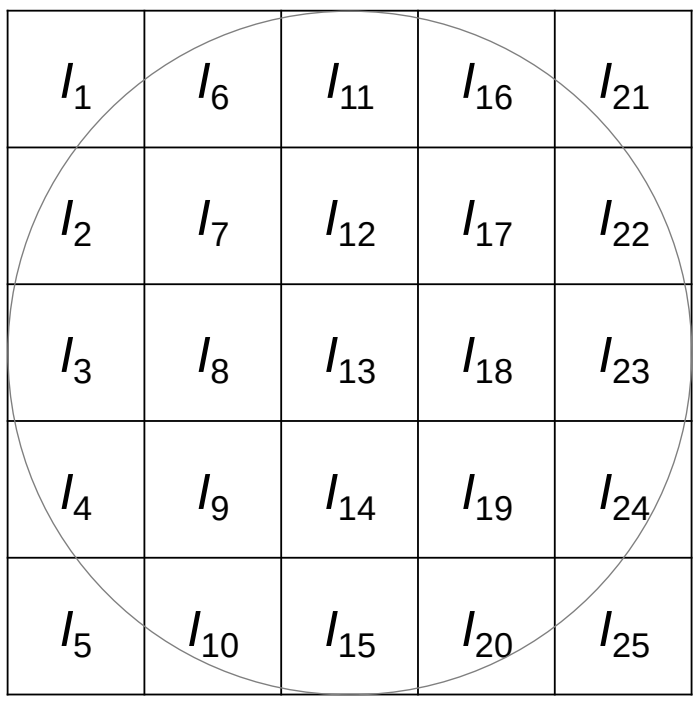

**Figure 2**. 5 × 5 bin locations

**Monte Carlo simulation**

The Monte Carlo simulation codes listed in [28] were used. A computer code to simulate the alterations in the electron trajectories due to the action of the Lorentz force was added [8, 18, 19]. We also added some programs to calculate the direction cosine of the electrons exiting the surface, so as to compute the intensity distributions of the electrons reaching a detector [29]. The configuration used in the simulations is illustrated in **Fig. 3**. The $x$–$y$ coordinate frame was located on the specimen surface. The following parameters were used: incident electron energy = 25 keV, number of incident electrons = $10^9$, inclination angle of the specimen from the horizontal = 70°, specimen atomic number = 26 (Fe), magnetic flux density = 2.2 T, direction of magnetisation vector = along $\pm x$ or $\pm y$ direction, and pattern centre (PC) position = [0.5, 0.4, 0.53]. The number of pixels in the square detector was 119 × 119. The definition of a PC position is described elsewhere [25]. The electrons with energies below 0.1 keV were not considered in the trajectory calculations.

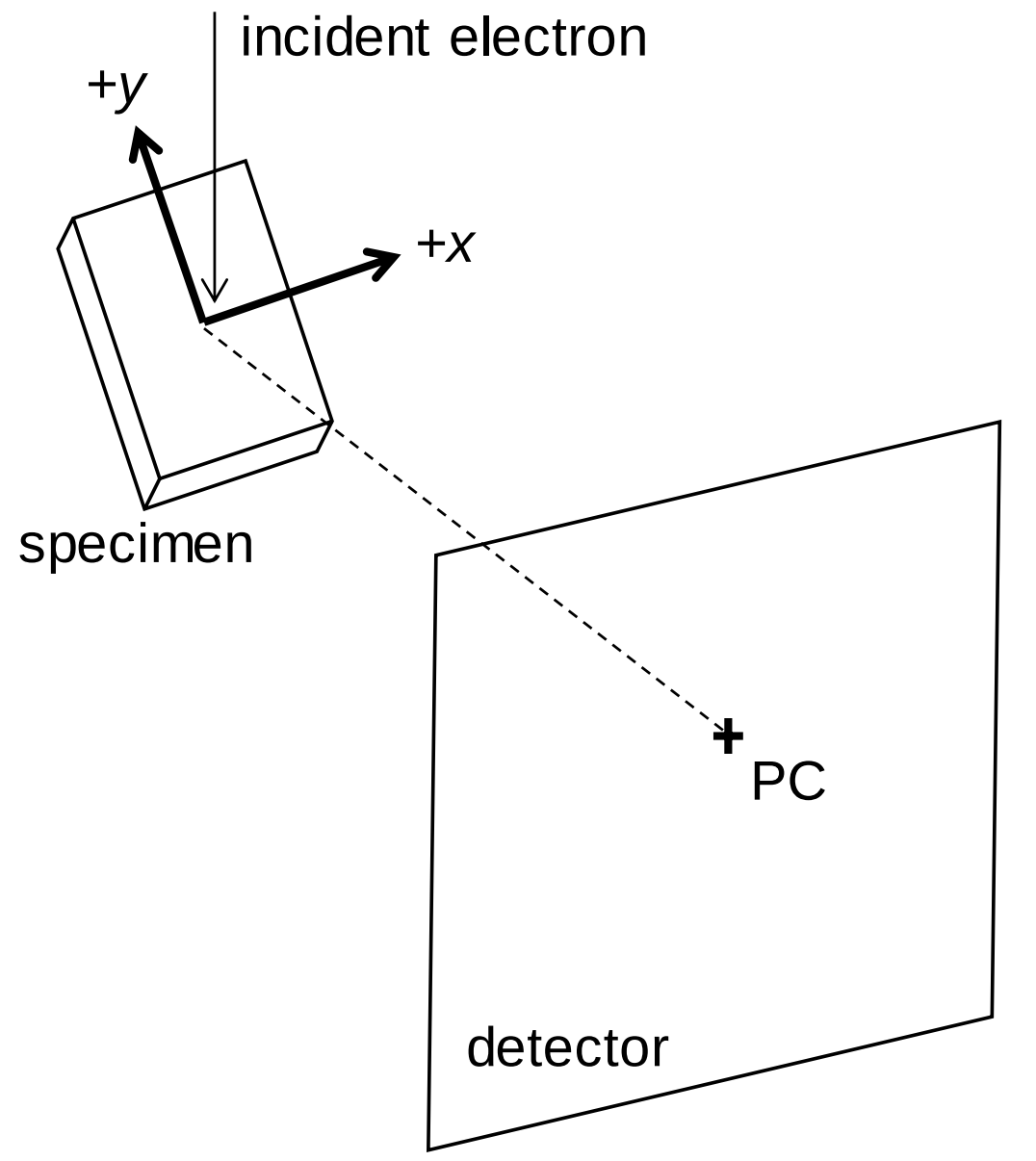

**Figure 3**. Configuration used in the simulation. PC denotes the pattern centre.

## Results

### Magnetic-domain imaging using an EBSD detector as an array of electron detectors

**Fig. 4** shows the inverse-pole-figure map and pole figure of the analysis area. The entire analysis area is monocrystalline, and the surface is nearly parallel to the (100) plane and perpendicular to the [001] plane. There are two easy magnetisation axes—[100] and [010] [30]—on the surface, which lie along the horizontal and vertical directions, as shown in **Fig. 4**(b). Thus, there are four possible in-plane magnetisation vectors that exist on the surface, namely, $B[100]$, $B[\bar{1}00]$, $B[010]$, and $B[0\bar{1}0]$.

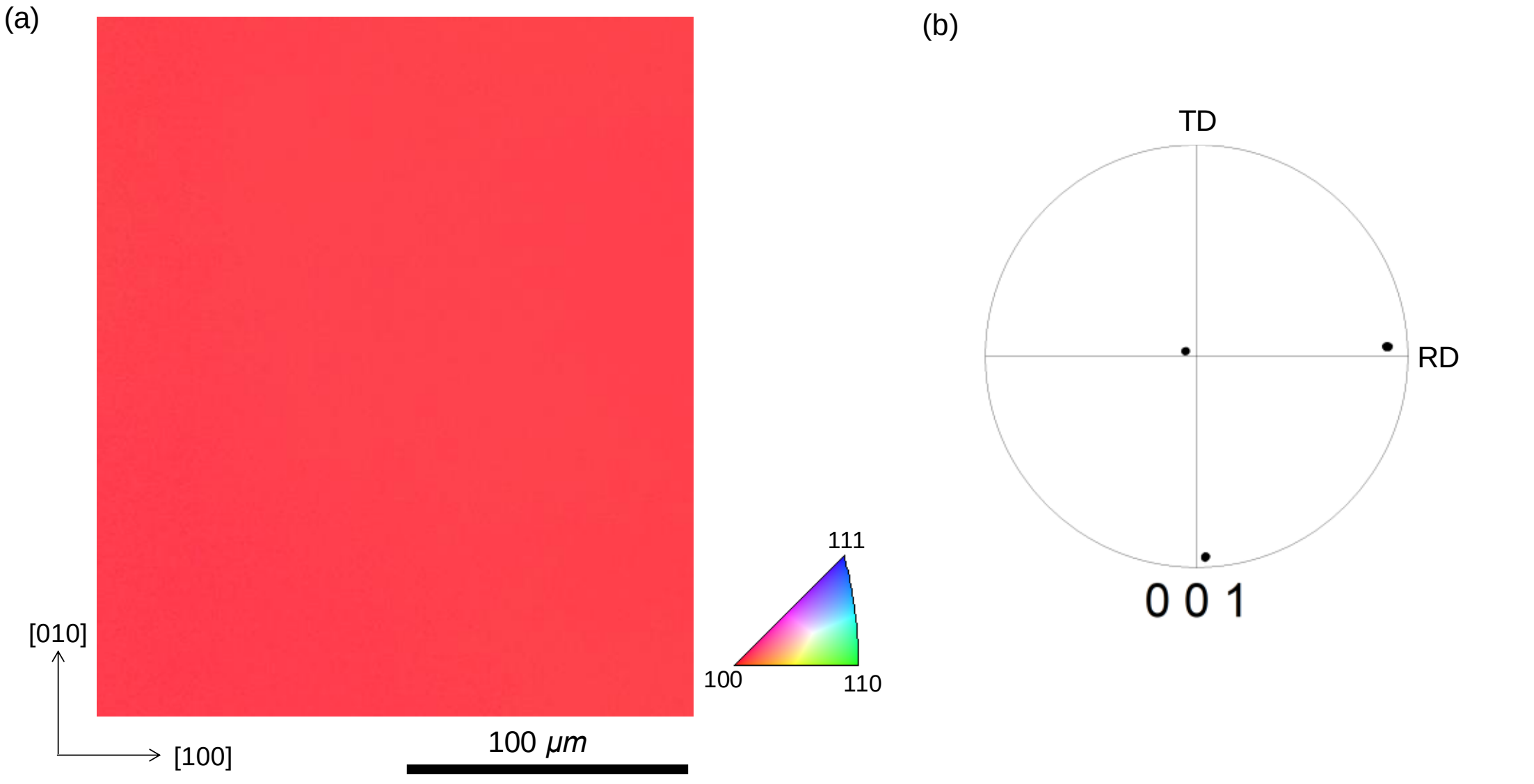

**Figure 4**. (a) Normal-direction inverse-pole-figure map and (b) [001] pole-figure map of the analysis area.

The scattered electron intensity map was constructed using the pixel intensities within the entire region of the EBSD pattern obtained from each analysis point in **Fig. 4**(a), and is displayed in **Fig. 5**. A fir-tree magnetic-domain pattern [9, 31] is clearly visible. In addition, the topographic contrast is superimposed on the magnetic-domain contrast, as a wavy intensity variation can be observed. This wavy nature of the surface morphology was caused by electropolishing.

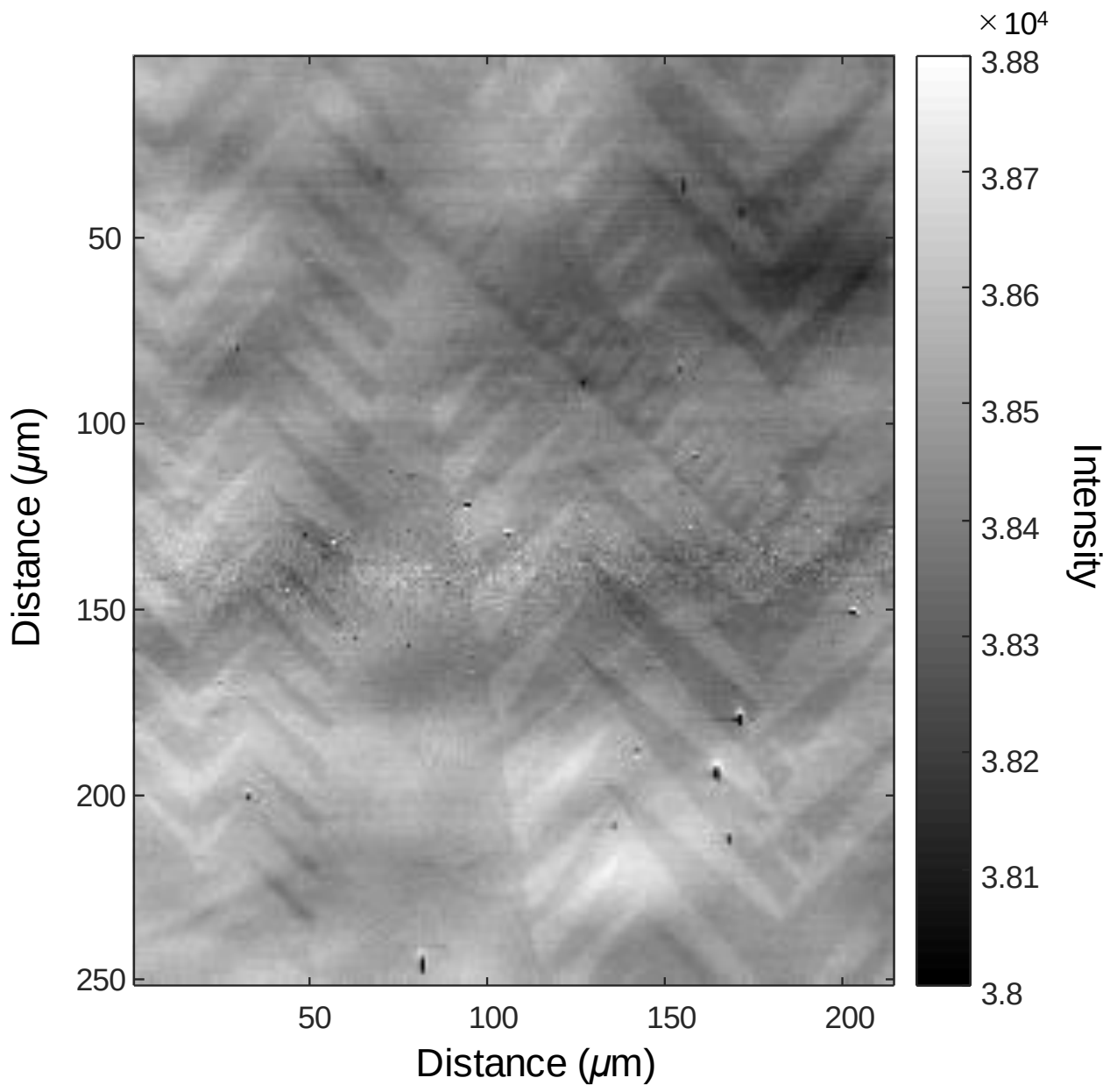

**Figure 5**. Electron image constructed using the scattered electrons captured over an entire region of the EBSD detector.

**Fig. 6**(a) displays the electron image constructed using the top region ($I_6 + I_{11} + I_{16}$) of the EBSD detector, showing only the topographic contrast. The magnetic contrast disappears because the backscattered electrons reaching the upper regions of the EBSD detector possess lower averaged kinetic energies than the electrons captured by the lower regions of the detector, under a typical WD setting (i.e., the position of the pattern centre (PC)) [29, 32, 33]. Because the type-II magnetic-domain contrast is caused by low-loss and elastically scattered electrons [8, 18], the electron image obtained using inelastically scattered electrons that reach the upper regions of an EBSD detector is unlikely to form a magnetic-domain contrast. Thus, it is possible to isolate or suppress the contrast of interest by selecting the location of the electron detector (i.e., by controlling the electron energy range used for imaging) [22, 34, 35].

**Fig. 6**(b) shows the electron image obtained using the bottom region ($I_{10} + I_{15} + I_{20}$) of the EBSD detector. The image looks identical to that in **Fig. 5,** except for the intensity range shown in the grey scale bar, which contains both the topographic and magnetic-domain contrasts. We can now suppress the topographic contrast through postprocessing. Because of the illumination effect, the topographic contrast can be reversed according to the virtual detector position, as shown in **Figs. 6**(a) and (b) (the intensities within the dashed black lines in these figures show opposite contrasts). The intensity variation caused by the topographical features can then be mitigated, to some extent, by adding the image intensities in **Fig. 6**(a) to those in **Fig. 6**(b).

Furthermore, an intensity variation is evident along the diagonal direction from the left bottom to the top right corner of **Figs. 5** and **6**(b). This intensity gradation is caused by the electron-beam scanning during the mapping analysis because the distance between an analysis point and the bin location changes constantly as the scan progresses, and is not negligibly small when a mapping area is scanned at low magnification. To mitigate this diagonal intensity gradation, an artificial background is created from the original electron image, which includes the diagonal intensity gradation, and is subtracted from the original electron image. The intensities of the four corners in the artificial background are set to the average intensities of 10 × 10 pixels located at the four corners in the electron image with a diagonal intensity gradation, and the intensities at the remaining points are determined through bilinear interpolation (**Fig. 6**(c)). Subsequently, image postprocessing is performed.

$$(I_{10} + I_{15} + I_{20}) - k_1\,(I_6 + I_{11} + I_{16}) - k_2\,I_{\mathrm{diagonal}}, \qquad (1)$$

where $k_i$ ($i$ = 1, 2) are the coefficients that change the image intensities to enhance the contrast of interest. $I_{\mathrm{diagonal}}$ corresponds to the image intensities shown in **Fig. 6**(c).

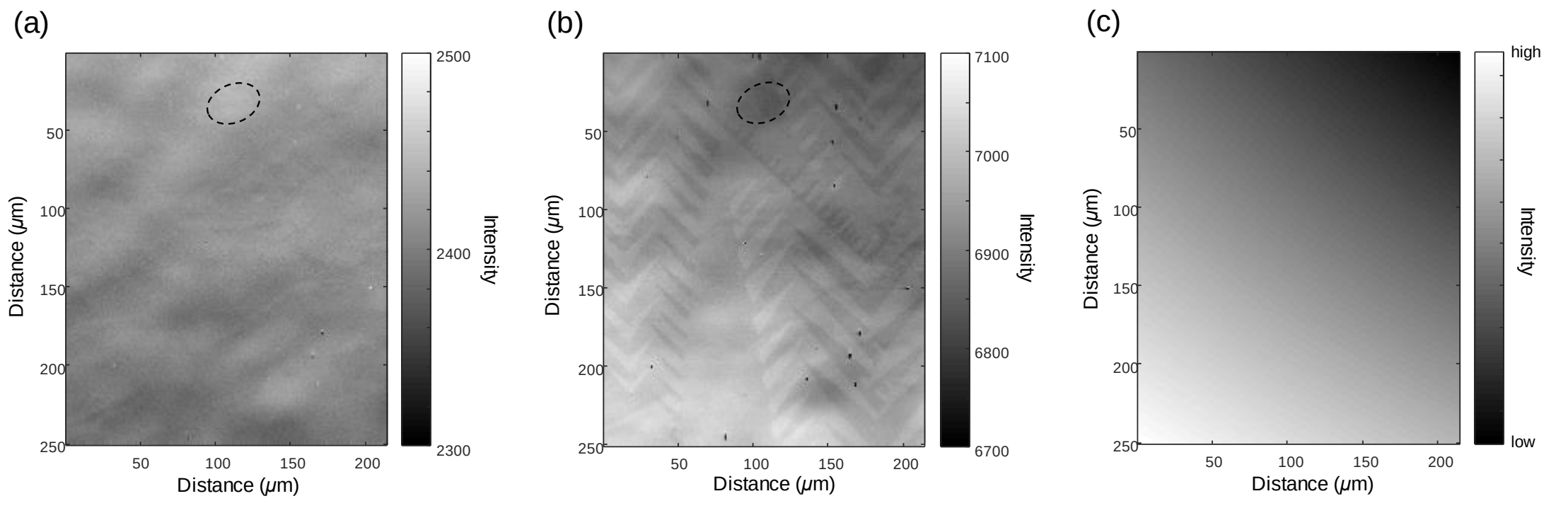

**Figure 6**. Electron images constructed using (a) $I_6 + I_{11} + I_{16}$ and (b) $I_{10} + I_{15} + I_{20}$. (c) Example of diagonal intensity gradation.

The postprocessed image is shown in **Fig. 7**, in which the magnetic-domain contrast is more evident, when compared to **Figs. 5** and **6**(b). It is possible to determine the direction of the magnetisation vectors in some magnetic domains. In the white-coloured (brightest) magnetic domains, where the backscattering coefficient is expected to be the highest, the magnetisation vector points leftward ($B[\bar{1}00]$), as shown in **Fig. 1**(a). In the darkest magnetic domain, the vector points rightward ($B[100]$). The directions of the magnetisation vectors in the grey-coloured magnetic domains should be either upward $B[010]$ or downward $B[0\bar{1}0]$. However, it is not possible to identify the difference from just the figures shown above, because the backscattering coefficients from the grey-coloured magnetic domains are almost identical.

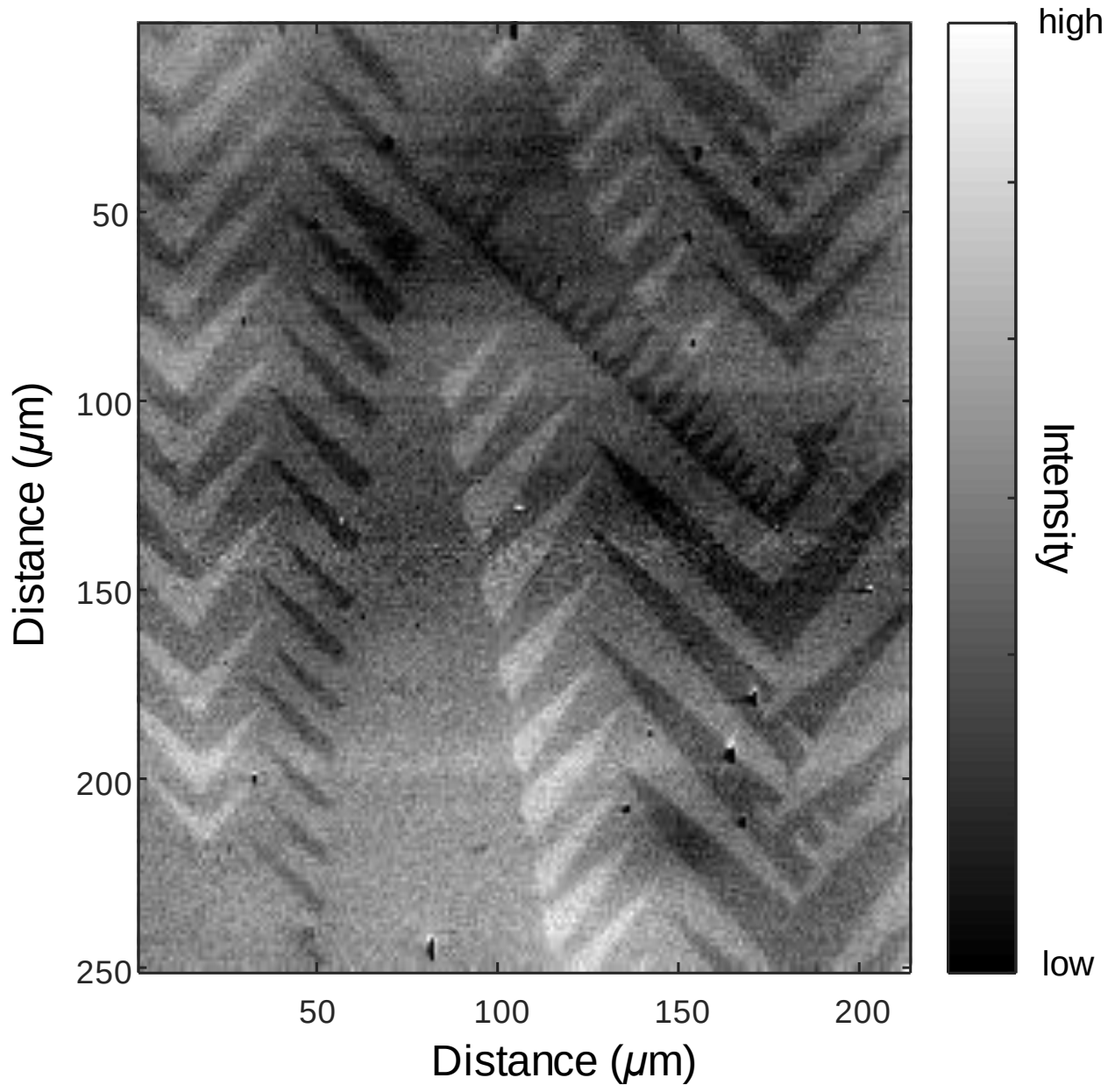

**Figure 7**. Postprocessed image.

To distinguish the magnetisation vectors ($B[010]$ or $B[0\bar{1}0]$), electron images were constructed using the left ($I_2 + I_3 + I_4$) and right ($I_{22} + I_{23} + I_{24}$) sides of the EBSD detector (**Fig. 8**). The postprocessed images are shown in Fig. 8. The grey-coloured magnetic domains in **Fig. 7** are now either white or dark. Therefore, a type-II magnetic-domain contrast can be obtained even in the case where the magnetisation vectors of opposite magnetisations are perpendicular to the tilt axis of the specimen. This implies that the anisotropy of electron scattering is influenced by the action of the Lorentz force, especially, along the lateral direction.

Based on the concepts presented in the Introduction section, the domain that is bright in the left portion of the detector and dark in the right is expected to possess a magnetisation vector pointing upward and vice versa. Then, all four possible in-plane magnetisation vectors are expected to be aligned, as shown in **Fig. 9**. This is verified through a sample rotation experiment and Monte Carlo simulations, as described in the following sections.

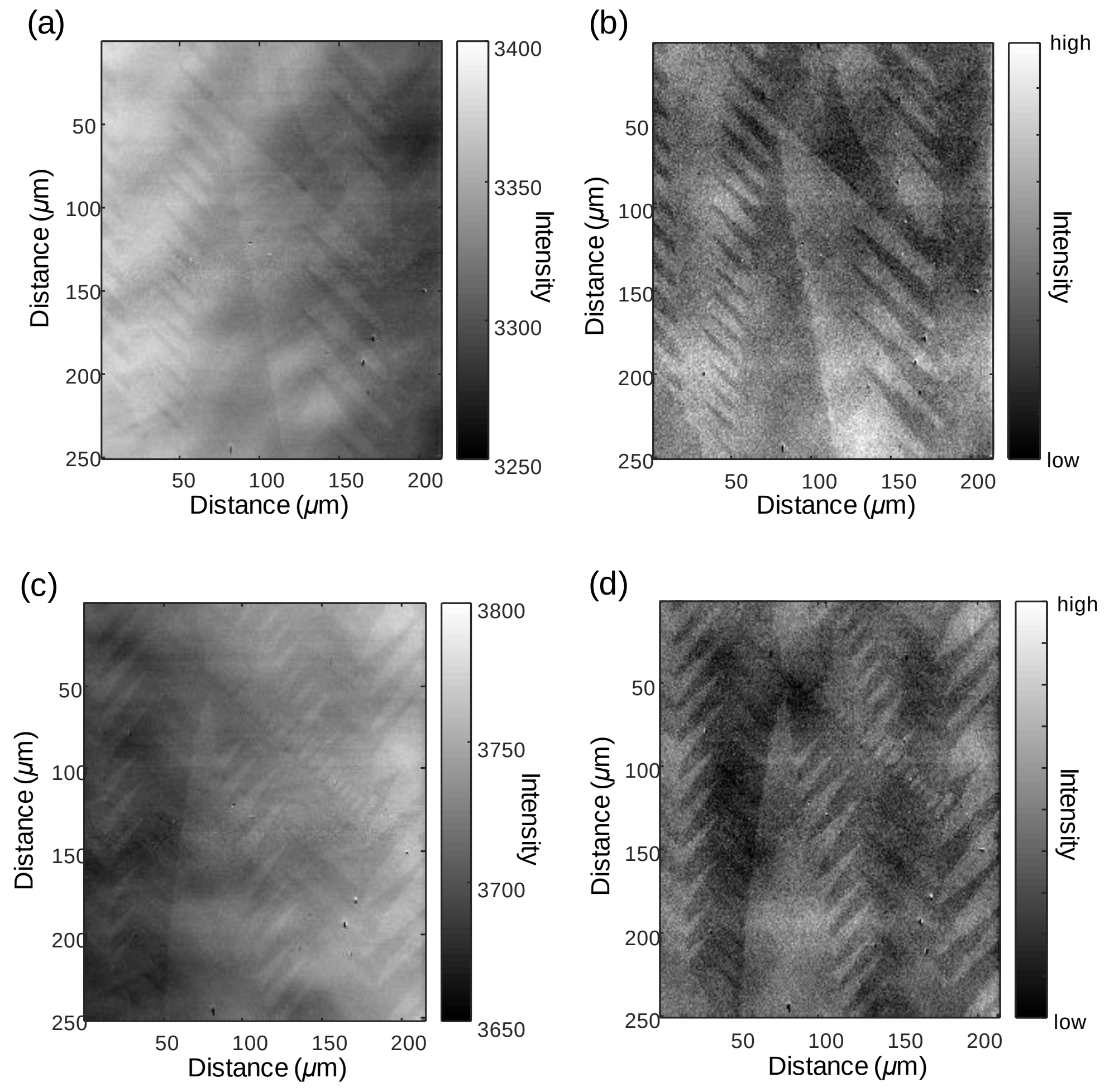

**Figure 8**. Electron images constructed using (a), (b) left ($I_2 + I_3 + I_4$) and (c), (d) right ($I_{22} + I_{23} + I_{24}$) sides of the EBSD detector. (a), (c) Raw images and postprocessed images (b) $(I_2 + I_3 + I_4) - k_1 (I_6 + I_{11} + I_{16}) - k_2 I_{\text{diagonal}}$, (d) $(I_{22} + I_{23} + I_{24}) - k_1 (I_6 + I_{11} + I_{16}) - k_2 I_{\text{diagonal}}$.

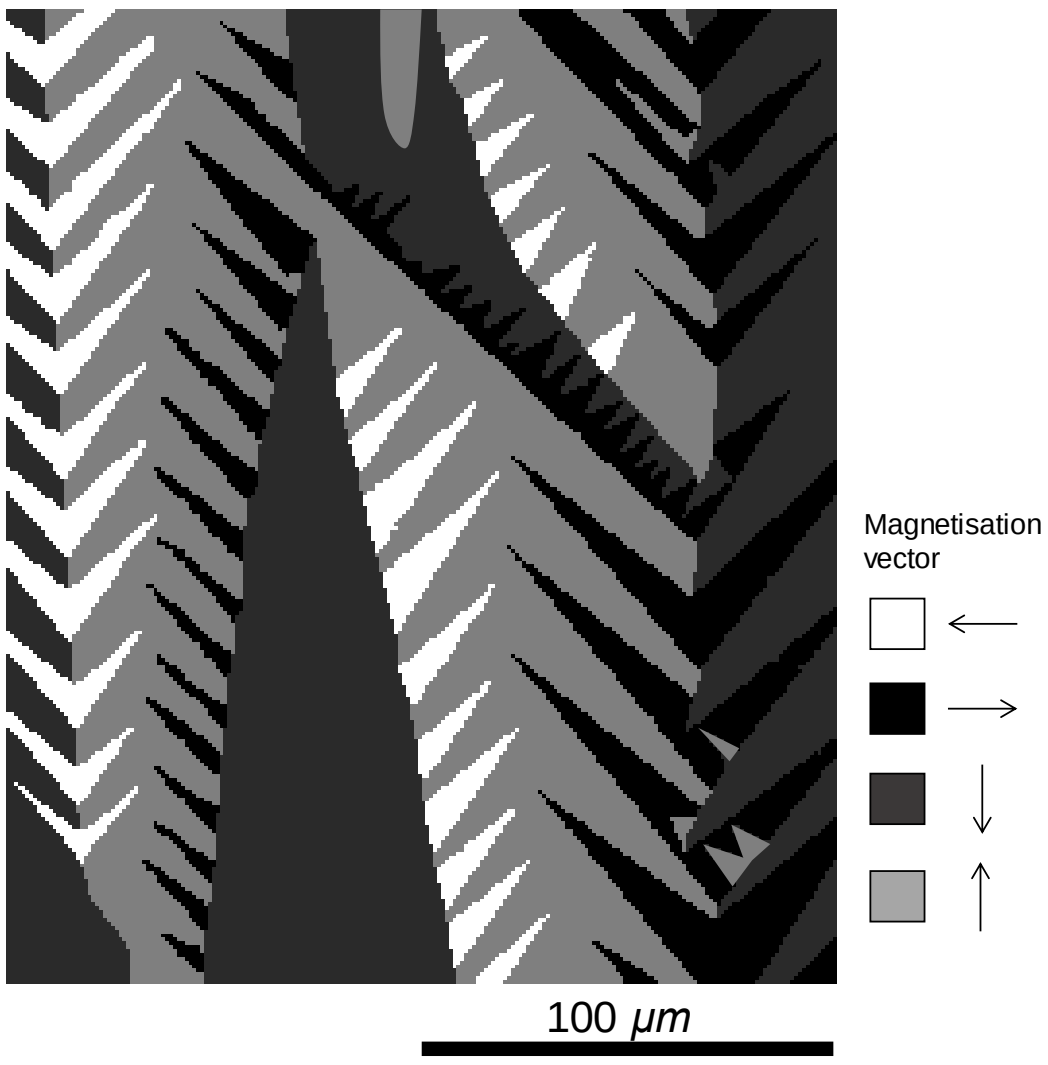

**Figure 9**. Directions of magnetisation vectors.

**Sample rotation experiment**

For experimental validation, the specimen was rotated about the surface normal by 90° and nearly the same area was scanned again by recording the EBSD patterns. The electron image constructed using the bottom portion of the EBSD detector ($I_{10} + I_{15} + I_{20}$) and the image postprocessed using *eq.* (1) are shown in **Fig. 10**. Some magnetic domains, shown in grey in **Fig. 7**, become the brightest (white), with magnetisation vectors pointing left. Thus, this magnetisation vector points in a downward direction after a −90° rotation. In this way, the directions of the magnetisation vectors in the grey-coloured magnetic domains in **Fig. 7** can be assigned, and there is no contradiction in the directions of the magnetisation vectors determined by the two approaches.

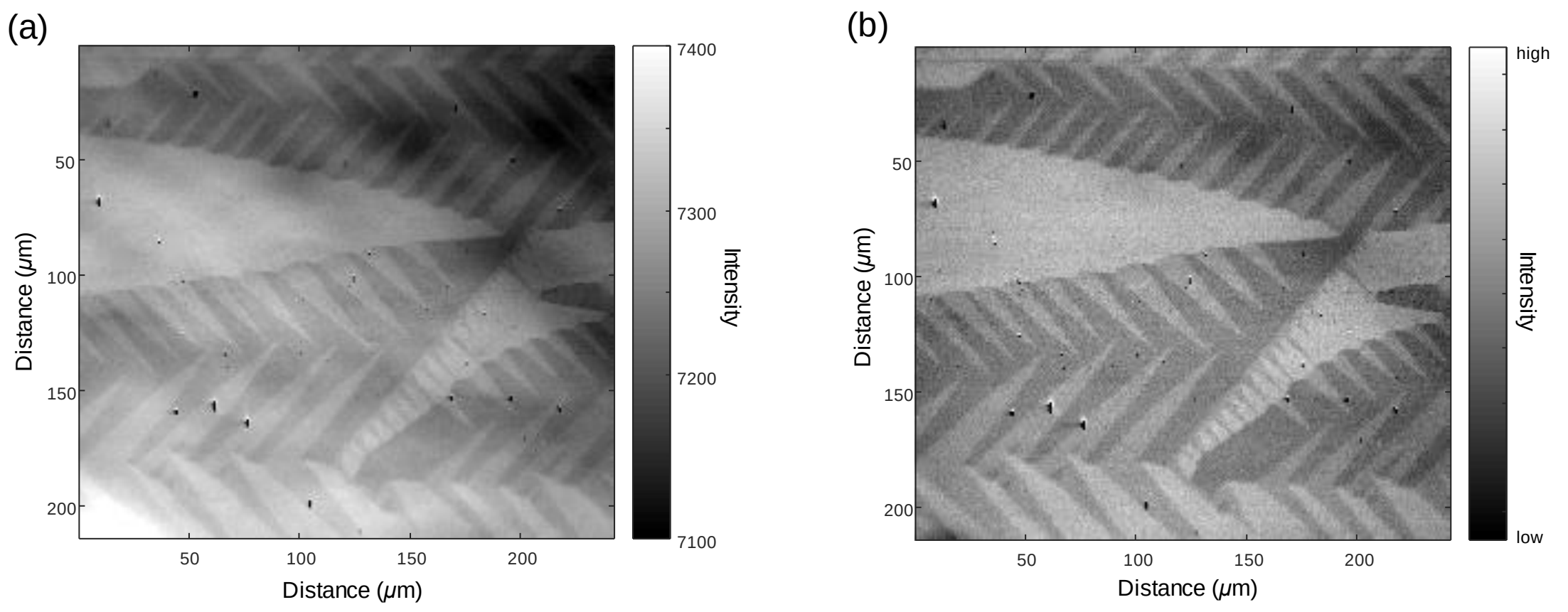

**Figure 10**. (a) Electron image obtained after 90° sample rotation about the surface normal, using the bottom region ($I_{10} + I_{15} + I_{20}$) as a virtual electron detector. (b) Image postprocessed using eq. (1).

**Monte Carlo simulation**

**Fig. 11** shows the backscattering coefficients from the magnetic domains with four different magnetisation vectors pointing in the $-x$, $+x$, $-y$, and $+y$ axis directions, respectively, obtained through Monte Carlo simulations. Nearly 60% of the incident electrons are backscattered, which agrees well with the previous experimental and simulation results [8, 36]. The highest coefficient is obtained from the domain with a magnetisation vector pointing toward $-x$ while the lowest is obtained from the domain with a vector pointing toward $+x$. The difference between the highest and lowest coefficients is merely 0.16%. These simulation results are in good agreement with that of the previous Monte Carlo simulation studies [8, 18, 19]. The coefficients from the magnetic domains with $\pm y$ vector directions show medium values and are nearly identical.

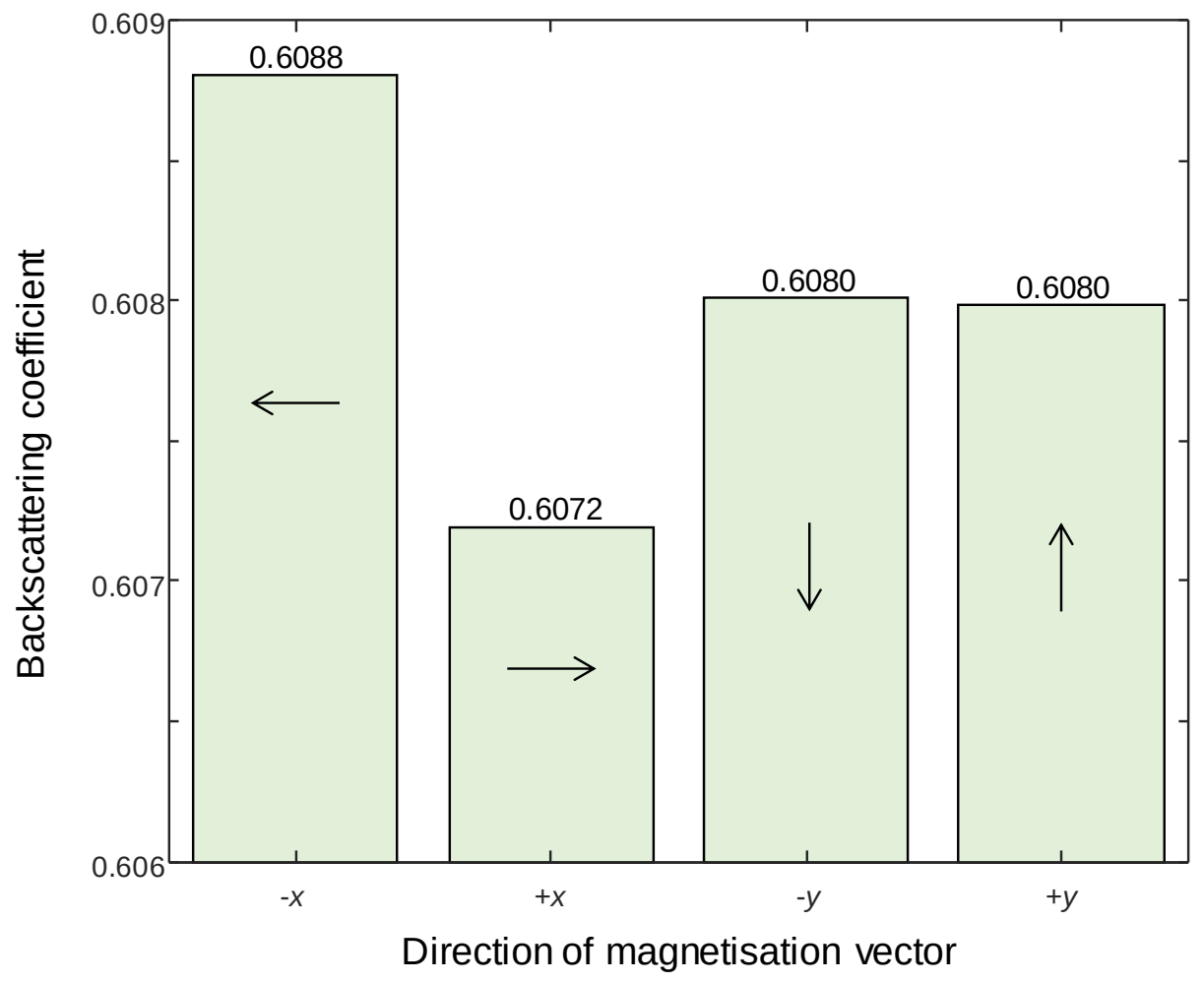

**Figure 11**. Monte Carlo simulation results of the backscattering coefficients from magnetic domains with four different in-plane magnetisation vectors.

**Figs. 12**(a) and (b) show the electron-intensity distributions on a detector obtained assuming $-x$ or $+x$ magnetisation vectors. The pixel-by-pixel intensity contrast is calculated as follows:

$$\text{Contrast}(i, j)\ (\%) = 100 \times (I_A(i, j) - I_B(i, j)) / (0.5 \times (I_A(i, j) + I_B(i, j))), \quad (2)$$

where $I_A(i, j)$ and $I_B(i, j)$ correspond to the intensity at pixel position $(i, j)$ ($i, j = 1, 2, \ldots$, number of pixels) in the electron images obtained from magnetic domains A and B, respectively. The calculated pixel-by-pixel contrast is shown in **Fig. 12(c)**. In the entire region of the detector, the contrast is beyond 0%, as the backscattering coefficient from the domain with the $-x$ magnetisation vector is higher. Long-range variations in contrast along the lateral direction are not observed (**Fig. 12**(d)). The contrast is apparent at the bottom of the detector, reaching 0.5–0.6% at the bottom edge (**Fig. 12**(e)). In the top region, the contrast is nearly 0%, indicating that magnetic-domain contrast does not exist when electrons are captured by the top portion of the detector. This explains why only the topographic contrast appears in **Fig. 6**(a) and why the magnetic-domain contrast is superimposed in **Fig. 6**(b).

**Figs. 13**(a) and (b) show the electron intensity distributions on a detector, obtained assuming $-y$ or $+y$ magnetisation vectors. Although the backscattering coefficients are almost identical between these magnetic domains, the pixel intensity contrast between the two images (**Fig. 13**(c)) shows variations along the lateral direction of the detector, as shown in **Fig. 13**(d). The magnetic domain with the $-y$ vector direction is darker (contrast $< 0$) on the left-hand side of the detector and brighter (contrast $>$ 0) on the right-hand side. This result is consistent with the idea presented in the Introduction section and explains the contrast in **Fig. 8**. The contrast at the sides of the detector reaches 0.2%, which is comparable to that in **Fig. 12**(d). This means that the magnetic-domain contrast caused by the backscattering-coefficient difference and that caused by the difference in the anisotropy of electron emissions are almost equal at the sides of the detector. It should be noted that the contrast variation along the lateral direction occurs while maintaining the total pixel intensities. Fig. 13(e) shows that the contrast does not vary along the vertical direction of the detector.

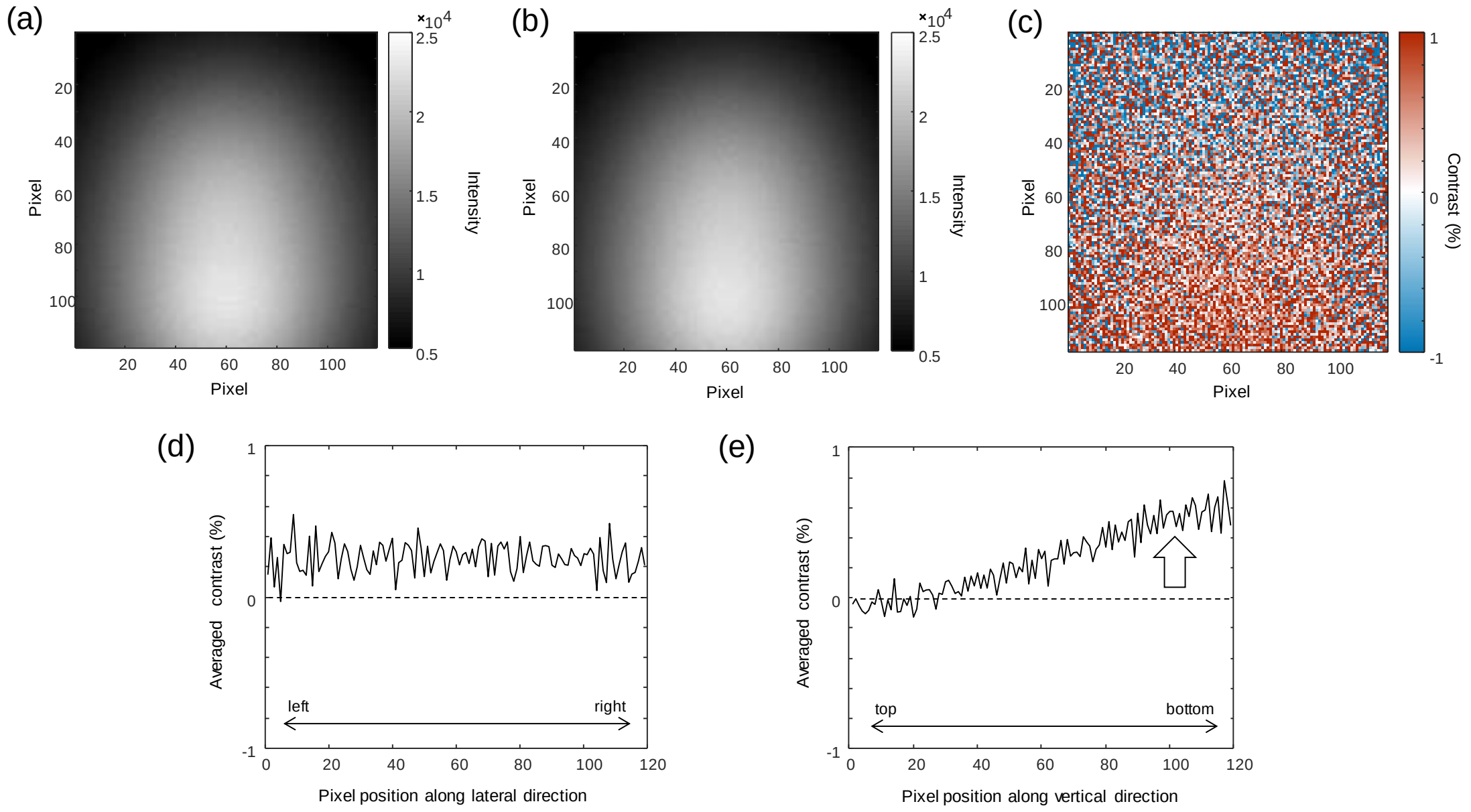

**Figure 12.** Electron-intensity distribution on a detector assuming specimen's magnetisation with magnetisation vectors along (a) $-x$ (leftward) and (b) $+x$ (rightward) directions. (c) Pixel-by-pixel contrast between the two patterns in (a) and (b). Averaged pixel intensity contrast in (c) along (d) lateral, and (e) vertical directions.

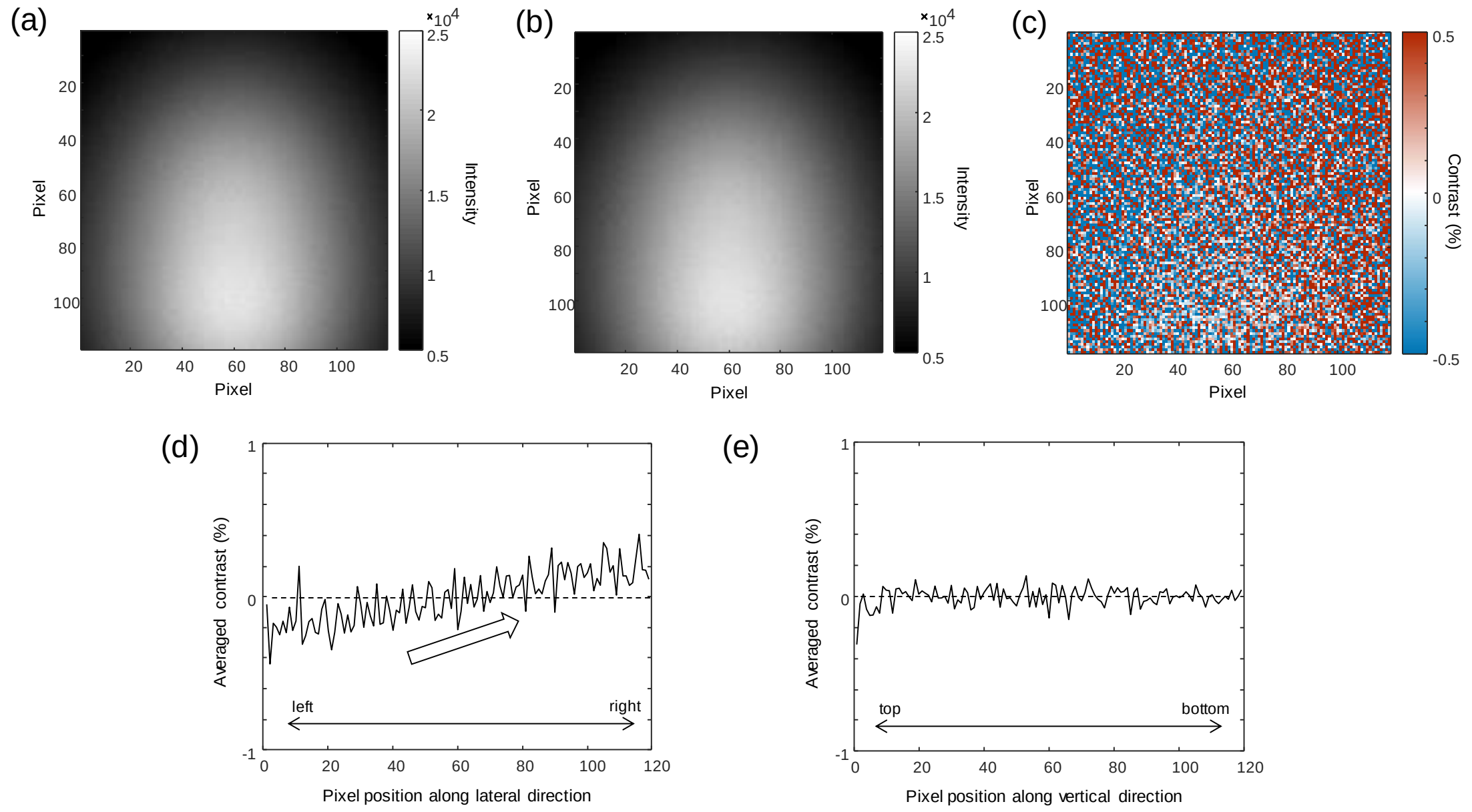

**Figure 13.** Electron intensity distribution on a detector assuming specimen's magnetisation with magnetisation vectors along (a) $-y$ (downward) and (b) $+y$ (upward) directions. (c) Pixel-by-pixel contrast between the two patterns in (a) and (b). Averaged pixel intensity contrast in (c) along (d) lateral and (e) vertical directions.

**Effect of internal magnetisation on the raw EBSD patterns**

Once all four magnetisation vectors are successfully assigned, the actual EBSD patterns obtained from opposite magnetisations can be compared. **Figs. 14**(a) and (b) show the EBSD patterns obtained from the $B[\bar{1}00]$ and $B[100]$ domains, respectively. The analysis points from which the two patterns are obtained are separated by 28 μm. The pixel-by-pixel intensity contrast between the two patterns shows that the contrast is evident in the bottom region of the EBSD pattern, while it is not clearly seen in the top region (**Fig. 14**(c)). The contrast is almost greater than zero in the entire region of the detector (**Figs. 14** (d) and (e)), indicating that the backscattering coefficient from the magnetic domain with the $B[\bar{1}00]$ vector is clearly larger than that from the domain with the $B[100]$ vector. These features are consistent with the Monte Carlo simulation results.

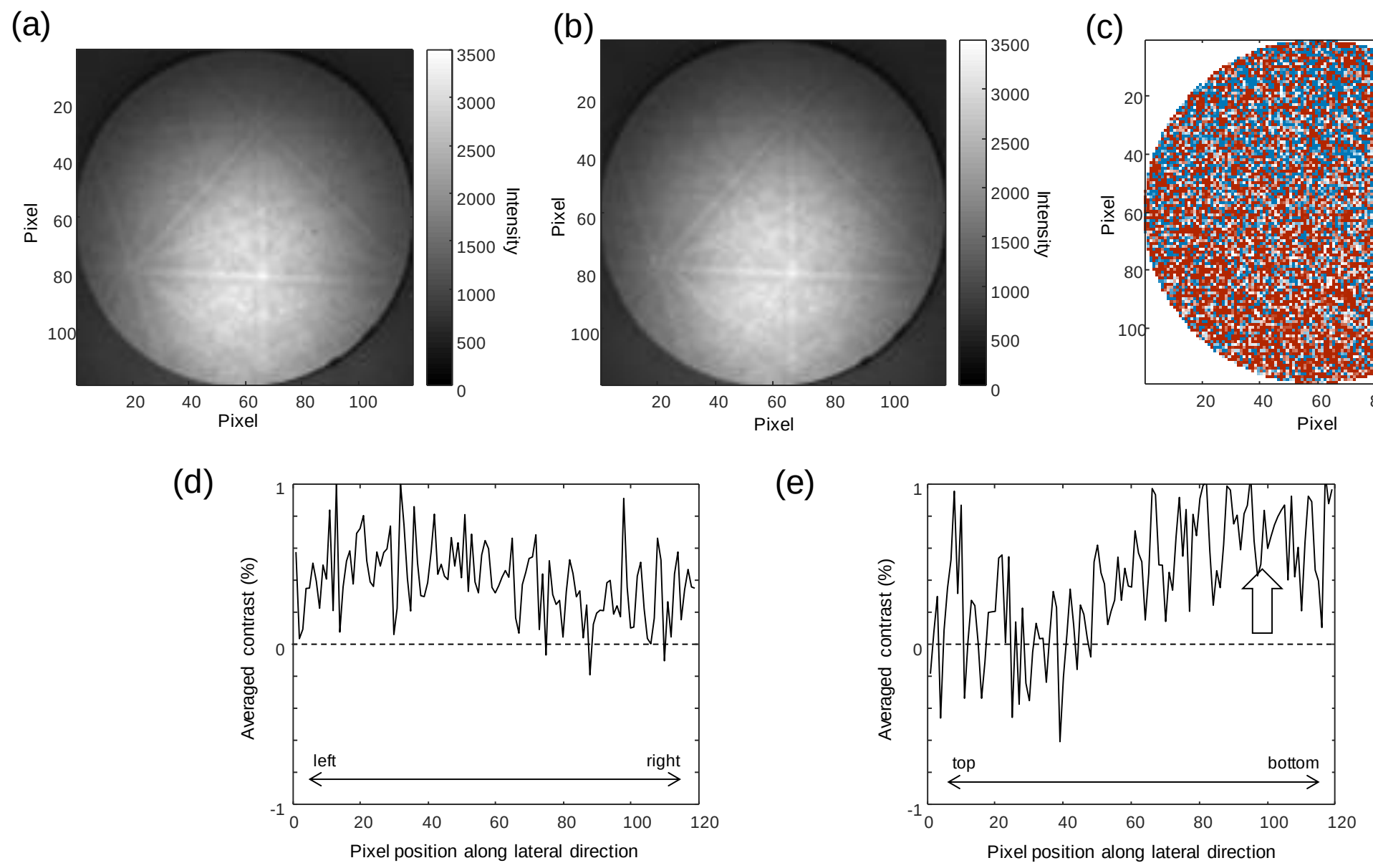

**Figure 14**. EBSD patterns obtained from domains with vectors along the (a) $B[\bar{1}00]$ (leftward) and (b) $B[100]$ (rightward) directions. (c) Pixel-by-pixel intensity contrast between the two EBSD patterns in (a) and (b). Averaged pixel intensity contrast in (c) along (d) lateral and (e) vertical directions.

**Figs. 15**(a) and (b) show the EBSD patterns obtained from the $B[010]$ and $B[0\bar{1}0]$ domains, respectively. The analysis points from which the two patterns are obtained are separated by 25 μm. The pixel-by-pixel intensity contrast between the two patterns is shown in **Fig. 15**(c). Again, the trend in the contrast variation along the lateral and vertical directions (**Figs. 15** (d) and (e)) corresponds well with the Monte Carlo simulation results. The difference between the experimental results and Monte Carlo simulation results is in the extent of the contrast. In **Fig. 13**(d), the contrast ranges nearly ±0.2%, whereas it is nearly ±0.4~0.5 % in **Fig. 15**(d). Winkelmann et al. [33] pointed out a problem with the use of the continuous slowing-down model in the Monte Carlo simulations of the quasi-elastic electron energy regime. In addition, the Monte Carlo simulations do not consider the diffraction/channelling phenomena. This might be related to the discrepancy in the contrasts between **Figs**. **13**(d) and **15**(d), although it is beyond the scope of this study to refine the electron trajectory simulation model for better quantitative analysis. Nevertheless, the Monte Carlo simulations used in this study explain the experimental tendency of the image-contrast variations for magnetic domains with different magnetisation vectors.

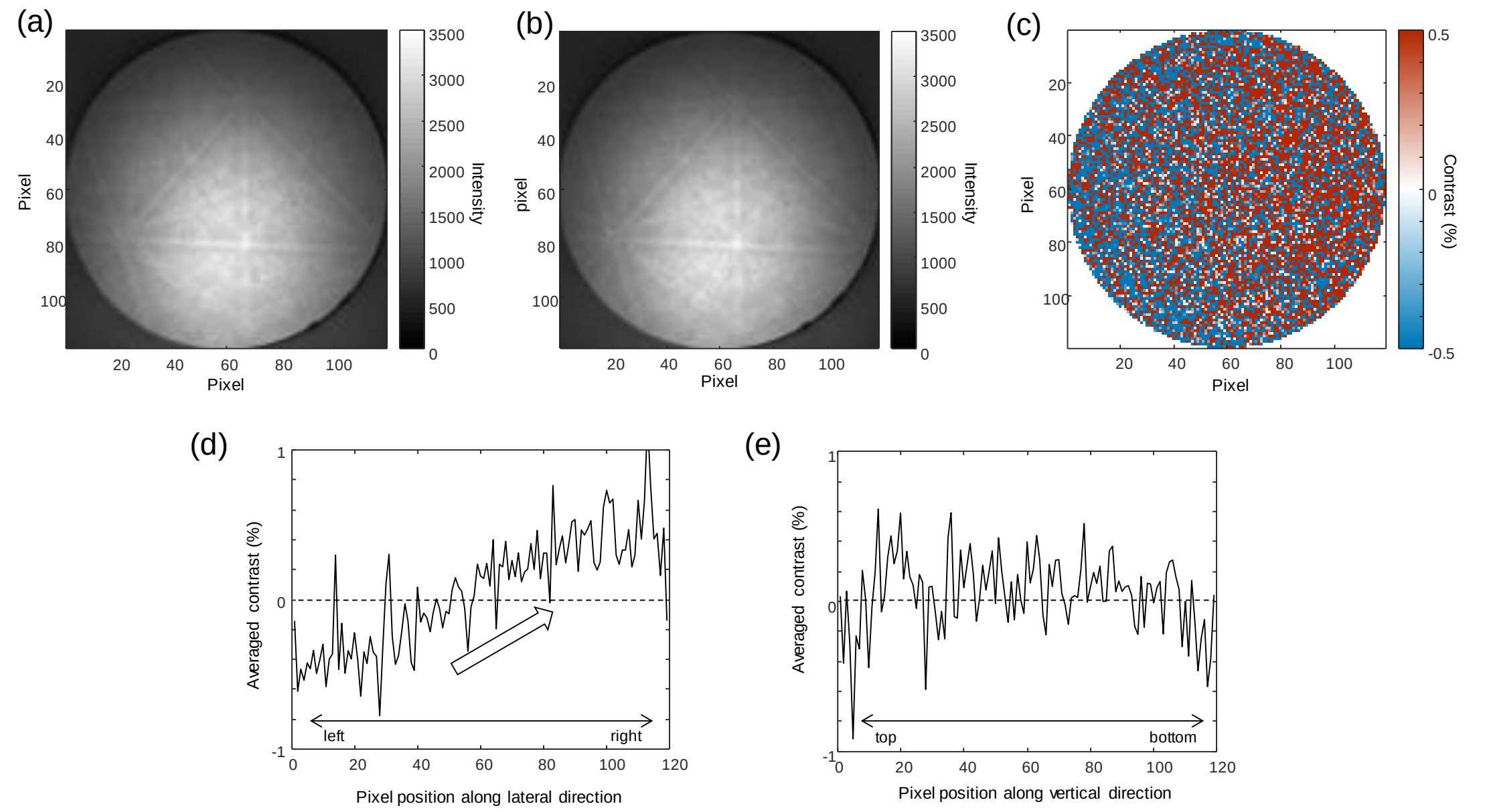

**Figure 15**. EBSD patterns obtained from domains with vectors along (a) $B[0\bar{1}0]$ (downward) and (b) $B[010]$ (upward) direction. (c) Pixel-by-pixel intensity contrast between the two EBSD patterns in (a) and (b). Averaged pixel intensity contrast in (c) along (d) lateral and (e) vertical directions.

It should be noted that the pixel intensity contrasts observed in **Figs. 14**(c) and **15**(c) are not evident along any Kikuchi band edge, indicating that the internal magnetisation in Fe does not alter the location of the Kikuchi bands at any recognisable level. The source point where diffracted electrons are generated might be shifted owing to the action of the Lorentz force, but the extent of the shift is expected to be quite small (i.e., in the order of the electron mean free path) and is not detectable. Indeed, the crystal orientation determined through the Hough transform is not altered by the presence of magnetic domains (see **Fig. 4**).

**Fig. 16**(a) summarises the contrasts in the electron images constructed by the electrons reaching $I_1$~$I_{25}$ bin positions, obtained in this study. It should be noted that the contrast variation over the bin positions is also dependent on the inclination angle, acceleration voltage of an SEM and crystal orientation. In this study, when using the top region, only the topographic contrast appears whereas the magnetic contrast is also superimposed in the middle and bottom regions of the detector. This contrast change is explained well by the electron take-off angle because the scattered electron-energy range is expected to be correlated with the take-off angle [22, 34, 35]. However, the electron take-off angle is not the sole factor determining the magnetic-domain contrast; the azimuthal angle (**Fig. 16**(b)) also needs to be considered [12], as the difference in the anisotropy of electron scattering, especially along the lateral direction, also contributes to the formation of the magnetic-domain contrasts at the side of the detector.

The electron take-off and azimuthal angles are dependent on PC location. A shorter camera length increases the azimuthal angle at the side of the detector. This might lead to a further increase in the magnetic-domain contrast caused by the difference in the electron emission anisotropy. The use of an energy-filtering system [10, 37] also increases the magnetic-domain contrast by selecting elastically scattered/low-loss electrons for imaging. Further improvements in the magnetic-domain contrast can be achieved based on the understanding of the contrast-formation mechanism, obtained in this study.

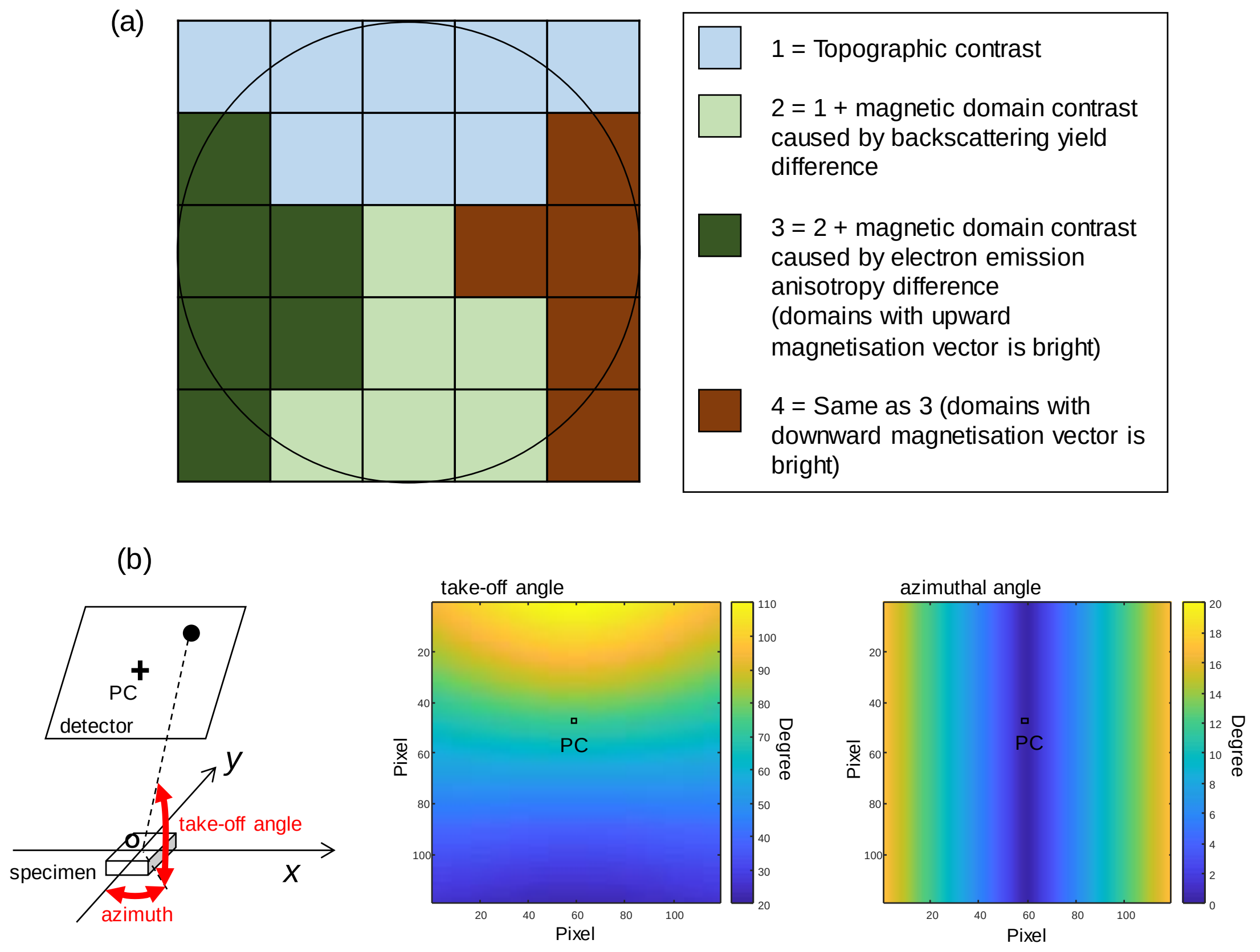

**Figure 16**. (a) Main cause of contrast in the electron images constructed by the electrons reaching $I_1$~$I_{25}$ bin positions. (b) Electron take-off and azimuthal angles as functions of the pixel position on a detector. The PC position is the same as that used in the Monte Carlo simulations.

## Conclusion

In this paper, we reported a new mechanism for the formation of type-II magnetic-domain contrasts caused by the differences in the anisotropy of electron emissions from the surface of Fe–Si steel produced by the Lorentz force. This was revealed using an EBSD detector as an array of electron detectors. The magnetic-domain contrast was quite small or unrecognisable when scattered electrons were captured by the upper portion of the detector, as inelastically scattered electrons were unlikely to contribute to the formation of a magnetic-domain contrast. The magnetic-domain contrast caused by the difference in backscattering yields was apparent when utilising the bottom centre region of the detector. In addition to this mechanism, an additional new mechanism, which was derived from the electron scattering anisotropy difference, also contributed to the formation of the magnetic-domain contrast when the electrons were captured by the side portion of the detector. The change in contrast according to the location of the virtual electron detectors allowed us to distinguish all four in-plane magnetisation vectors on the Fe–Si (100) surface without sample rotation.

**References**

1. “Advanced scanning electron microscopy and X-ray microanalysis”, D.E. Newbury, D.C. Joy, P. Echlin, C.E. Fiori, J.I. Goldstein (Eds.), Springer (1986)

2. “Electron backscatter diffraction in materials science”, A.J. Schwartz, M. Kumar, B.L. Adams, D.P. Field (Eds.), Springer (2009)

3. J.R. Dorsey, *1st Nat. Conf: on Electron Probe Microanalysis,* Maryland, U.S.A. (1966)

4. J. Philibert, R. Tixier, “Effets de contraste cristallin en microscopie electronique a balayage” *Micron* 1 (1969) 174-186 (in French)

5. J.R. Banbury, W.C. Nixon, “The direct observation of domain structure and magnetic fields in the scanning electron microscope”, *J. Sci. Instrum.*, 44 (1967) 889-892

6. J. Unguris, G. Hembree, R.J. Celotta, D.T. Pierce, “Investigations of magnetic microstructures using scanning electron microscopy with spin polarisation analysis”, *J. Magn. Magn. Mater.*, 54-57 (1986) 1629-1630

7. H. Akamine, S. Okumura, S. Farjami, Y. Murakami, M. Nishida, “Imaging of surface spin textures on bulk crystals by scanning electron microscopy”, *Sci. Rep.*, 22 (2016) 37265

8. D.J. Fathers, J.P. Jakubovics, D.C. Joy, D.E. Newbury, H. Yakowitz, “A new method of observing magnetic domains by scanning electron microscopy I. Theory of the image contrast”, *Phys. Stat. Sol. (a)* 20 (1973) 535-544

9. D.J. Fathers, J.P. Jakubovics, D.C. Joy, D.E. Newbury, H. Yakowitz, “A new method of observing magnetic domains by scanning electron microscopy II. Experimental confirmation of the theory of the image contrast”, *Phys. Stat. Sol. (a)* 22 (1974) 609-619

10. T. Yamamoto, H. Nishizawa, K. Tsuno, “Magnetic domain contrast in backscattered electron images obtained with a scanning electron microscope”, *Phil. Mag.* 34 (1976) 311-325

11. R. Shimizu, T. Ikuta, M. Kinoshita, T. Murayama, H. Nishizawa, T. Yamamoto, “High contrast observation of magnetic domain with high voltage SEM”, *Jpn. J. Appl. Phys.*, 15 (1976) 967-981

12. N. Brodusch, H. Demers, R. Gauvin, "Imaging with a commercial electron backscatter diffraction (EBSD) camera in a scanning electron microscope: A review", *J. Imaging* 4 (2018) 88

13. T. Ickler, H. Meckbach, F. Ziesmann, A. Bruckner-Foit, "Assessing the influence of crystallographic orientation, stress and local deformation on magnetic domains using electron backscatter diffraction and forescatter electron imaging" *Ultramicroscopy* 198 (2019) 33-42

14. M. Gallaugher, N. Brodusch, R. Gauvin, R.R. Chromik, "Magnetic domain structure and crystallographic orientation of electrical steels revealed by a forescatter detector and electron backscatter diffraction" *Ultramicroscopy* 142 (2014) 40-49

15. A. Nadoum, F. Robinson, S. Birosca, "On the correlation between magnetic domain and crystallographic grain orientation in grain oriented electrical steels" *J. Magn. Magn. Mater.*, 494 (2020) 165772

16. T. Kohashi, M. Konoto, K. Koike, "High-resolution spin-polarized scanning electron microscopy (spin SEM)", *J. Electron Microsc.*, 59 (2010) 43-52

17. T. Kohashi, K. Koike, "A spin-polarized scanning electron microscope with 5 nm-resolution", *Jpn. J. Appl. Phys.*, 40 (2001) L11264-L1266

18. D.E. Newbury, H. Yakowitz, R.L. Myklebust, "Monte Carlo calculations of magnetic contrast from cubic materials in the scanning electron microscope", *Appl. Phys. Lett.*, 23 (1973) 488-490

19. T. Ikuta, R. Shimizu, "Magnetic domain contrast from ferromagnetic materials in the scanning electron microscope", *Phys. Stat. Sol. (a)* 23 (1974) 605-613

20. O.C. Wells, L.M. Gignac, C.E. Murray, A. Frye, J. Bruley, "Use of backscattered electron detector arrays for forming backscattered electron images in the scanning electron microscope" *Scanning* 28 (2006) 27–31.

21. R.A. Schwarzer, J. Hjelen, "Backscattered electron imaging with an EBSD detector", *Microsc. Today* 23 (2015) 12-17

22. S.I. Wright, M.M. Nowell, R. Kloe, P. Camus, T. Rampton, "Electron imaging with an EBSD

detector", *Ultramicroscopy* 148 (2015) 132-145

23. S. Taguchi, A. Sakakura, T. Yasunari, "Method for producing silicon steel strips having cube-on-face orientation" US patent 3163564 (1964)

24. S. Arai, M. Mizokami, M. Yabumoto, "The magnetic properties of doubly oriented 3% Si-Fe and its application" *J. Appl. Phys.*, 67 (1990) 5577-5579

25. T. Tanaka, A.J. Wilkinson, "Pattern matching analysis of electron backscatter diffraction patterns for pattern centre, crystal orientation and absolute elastic strain determination – accuracy and precision assessment", *Ultramicroscopy* 202 (2019) 87-99

26. T. Tanaka, N. Maruyama, N. Nakamura. A.J. Wilkinson, "Tetragonality of Fe-C martensite – a pattern matching electron backscatter diffraction analysis compared to X-ray diffraction", *Acta Mater.*, 195 (2020) 728-738

27. A. Winkelmann, C.T. Cowan, F. Sweeney, A.P. Day, P. Parbrook, "Many-beam dynamical simulation of electron backscatter diffraction patterns", *Ultramicroscopy* 107 (2007) 414-421

28. L. Reimer, D. Stelter, "FORTRAN 77 monte carlo program for minicomputers using mott cross-sections", *Scanning* 8 (1986) 265-277

29. P.G. Callahan, M.D. Graef, "Dynamical electron backscatter diffraction patterns. Part I: Pattern simulations", *Microsc. Microanal.*, 19 (2013) 1255-1265

30. K. Honda, S. Kaya, *Sci. Rep. Tohoku Univ*. 15 (1926) 721

31. H.J. Williams, R.M. Bozorth, W. Shockley, "Magnetic domain patterns on single crystals of silicon iron", *Phys. Rev.*, 75 (1949) 155-178

32. F. Ram, M.D. Graef, "Energy dependence of the spatial distribution of inelastically scattered electrons in backscatter electron diffraction", *Phys. Rev. B* 97 (2018) 134104

33. A. Winkelmann, T.B. Britton, G. Nolze, "Constraints on the effective electron energy spectrum in backscatter Kikuchi diffraction", *Phys. Rev. B* 99 (2019) 064115

34. L. Reimer “Scanning electron microscopy”, Springer-Verlag, Berlin Heidelberg, (1998).

35. T. Aoyama, M. Nagoshi, H. Nagano, K. Sato, S. Tachibana, “Selective backscattered electron imaging of material and channeling contrast in microstructures of scale on low carbon steel controlled by accelerating voltage and take-off angle” *ISIJ Int*., 51 (2011) 1487

36. G. Neubert, S. Rogaschewski, “Backscattering coefficient measurements of 15 to 60 keV electrons for solids at various angles of incidents”, *Phys. Stat. Sol. (a)* 59 (1980) 35-41

37. A. Deal, T. Hooghan, A. Eades, “Energy-filtered electron backscatter diffraction”, *Ultramicroscopy* 108 (2008) 116-125

## Appendix

Electron images constructed by the electrons reaching the $I_1$~$I_{25}$ bins are listed below.

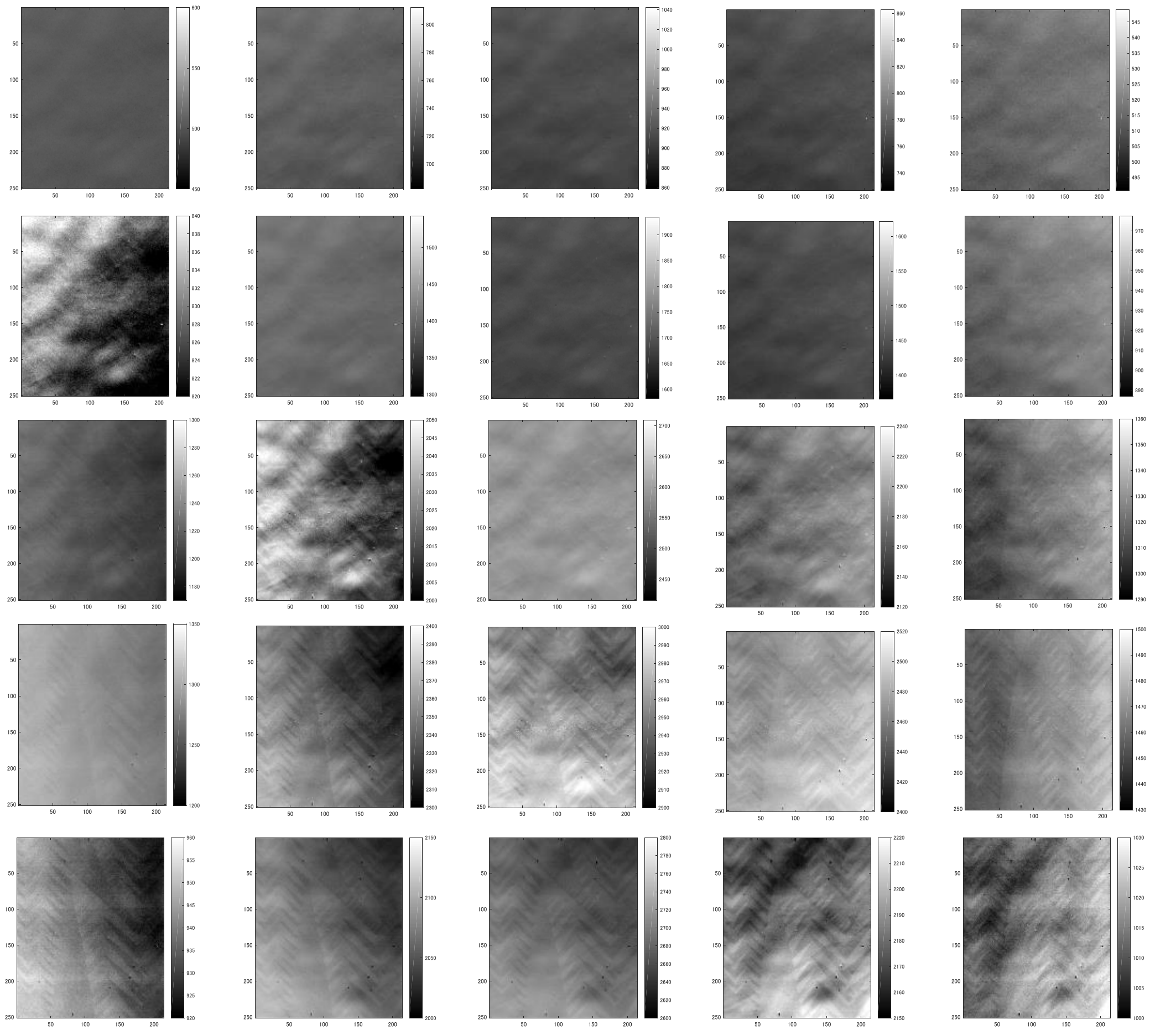

**Figure A-1**. Electron images constructed by the electrons reaching $I_1$~$I_{25}$ bins.